\documentclass[a4paper,fleqn,usenatbib]{mnras}

\usepackage{newtxtext,newtxmath}
\usepackage[T1]{fontenc}
\usepackage{ae,aecompl}

\usepackage{graphicx}	% Including figure files
\usepackage{amsmath}	% Advanced maths commands
\usepackage{amssymb}	% Extra maths symbols
\usepackage{graphicx}	% Including figure files

\title[]{High spectral resolution observations of HNC$_3$ and HCCNC in the L1544 prestellar core}%\thanks{Based on observations carried out with the IRAM 30 m Telescope. IRAM is supported by INSU/CNRS (France), MPG (Germany), and IGN (Spain).}}

\author[C. Vastel et al.]{\Large{C. Vastel$^{1}$\thanks{E-mail: cvastel@irap.omp.eu}, K. Kawaguchi$^{2}$, D. Qu\'enard$^{3}$,
M. Ohishi$^{4}$, B. Lefloch$^{5}$, R. Bachiller$^{6}$, H.~S.~P. M{\"u}ller$^{7}$\thanks{E-mail: hspm@ph1.uni-koeln.de}}
\\
$^{1}$IRAP, Universit\'e de Toulouse, CNRS, UPS, CNES, Toulouse, France\\
$^{2}$Graduate School of National Science and Technology, Okayama University, 3-1-1 Tsushima-naka, Okayama 700-8530, Japan\\
$^{3}$School of Physics and Astronomy, Queen Mary University of London, Mile End Road, London E1 4NS, UK\\
$^{4}$Astronomy Data Center, National Astronomical Observatory of Japan, 2-21-1 Osawa, Mitaka, Tokyo 181-8588, Japan\\
$^{5}$Universit\'e de Grenoble Alpes, CNRS, IPAG, 38000 Grenoble, France\\
$^{6}$Observatorio Astron\'omico Nacional (OAN, IGN), Calle Alfonso XII, 3, 28014 Madrid, Spain\\
$^{7}$I.~Physikalisches Institut, Universit{\"a}t zu K{\"o}ln, Z{\"u}lpicher Str. 77, 50937 K{\"o}ln, Germany
}

\date{Accepted 2017 November 27. Received 2017 November 21; in original form 2017 October 21}

\pubyear{2015}

\begin{document}
\label{firstpage}
\pagerange{\pageref{firstpage}--\pageref{lastpage}}
\maketitle

\begin{abstract}
HCCNC and HNC$_3$ are less commonly found isomers of cyanoacetylene, HC$_3$N, a molecule that is widely found in diverse 
    astronomical sources. We want to know if HNC$_3$ is present in sources other than the dark cloud TMC-1 and how 
    its abundance is relative to that of related molecules. We used the ASAI unbiased spectral survey at IRAM 30m towards the prototypical 
    prestellar core L1544 to search for HNC$_3$ and HCCNC which are by-product of the HC$_3$NH$^+$ recombination, previously detected in this source. We performed a combined analysis 
    of published HNC$_3$ microwave rest frequencies with thus far unpublished millimeter 
    data because of issues with available rest frequency predictions. We determined new spectroscopic parameters for HNC$_3$, produced new predictions and detected 
    it towards L1544. We used a gas-grain chemical modelling to predict the abundances of N-species and compare with the observations. The modelled abundances 
   are consistent with the observations, considering a late stage of the evolution of the prestellar core. However the calculated abundance of HNC$_3$ was found 5--10 times higher than the observed one. The HC$_3$N, HNC$_3$ and HCCNC versus HC$_3$NH$^+$ ratios are compared in the TMC-1 dark cloud and the L1544 prestellar core.
\end{abstract}

\begin{keywords}
astrochemistry---line: identification---ISM: abundances---ISM: molecules---ISM: individual objects (L1544)
\end{keywords}

%%%%%%%%%%%%%%%%%%%%%%%%%%%%%%%%%%%%%%%%%%%%%%%%%%

%%%%%%%%%%%%%%%%% BODY OF PAPER %%%%%%%%%%%%%%%%%%
\section{Introduction}

Given their powerful diagnostic ability, several unbiased spectral surveys have been obtained 
in the past few years in the millimeter and sub-millimeter bands accessible from the ground obtained in the direction of star forming regions \citep[e.g.][]{herbst2009}.  
By far the most targeted sources are hot cores, the regions of high mass protostars formation, 
where the dust temperature exceeds the sublimation temperature of the water-ice grain mantles, 
$\sim$100~K. So far, unbiased spectral surveys of low-mass solar type objects on a wide frequency range have been 
carried out in IRAS 16293-2422 \citep{blake1994,caux2011,jorgensen2016}. 
The increased capabilities of the IRAM 30~m telescope make access to a new level in the 
investigation of molecular complexity in star-forming regions with the ASAI (Astrochemical 
Studies At IRAM) Large program (see Lefloch et al. 2017 for a review) in the millimeter regime (project number 012-12). 
One of the targeted sources is the prototypical prestellar core, L1544, located in the Taurus 
molecular cloud complex (d $\sim$140~pc). This source is characterised by a high central density 
($\ge$ 10$^6$ cm$^{-3}$), a low temperature ($\le$ 10~K), a high CO depletion and a large degree 
of molecular deuteration \citep{crapsi2005,vastel2006,spezzano2013}. 
The study of the many tracers detected in this source led to the reconstruction of its physical 
and dynamical structure \citep{keto2014}. This core is also famous for being the first 
prestellar core where water has been detected \citep{caselli2012} and its line profile, 
characteristic of gravitational contraction, is confirming that L1544 is on the verge of the collapse. The ASAI high spectral resolution and high sensitivity (average rms of $\sim$4mK in a 50~kHz 
frequency bin) led to many discoveries. For example, we have been able to obtain a full census 
of oxygen-bearing iCOMs (interstellar complex organic molecules), and it was found 
that a non-thermal desorption mechanism is possibly responsible for the observed emission of 
methanol and COMs from the same external layer \citep{vastel2014}. 
The detection of the cyanomethyl radical (CH$_2$CN) for the first time in a prestellar core was 
reported by \citet{vastel2015b} who were able to identify resolved hyperfine transitions 
of the \textit{ortho} and \textit{para} forms from the calculated frequencies. 
Finally, protonated cyanides, such as HCNH$^+$ and HC$_3$NH$^+$, have also been 
detected for the first time in a prestellar core \citep{quenard2017}. 
The high spectral resolution of the observations allows to resolve the hyperfine structure 
of HCNH$^+$. The study of some nitrogen species linked to their production (HCN, HNC, HC$_3$N) 
coupled with a gas-grain chemical modelling shows that the emission of these ions originates 
in the external layer where non-thermal desorption of other species was previously observed. 

In the present study we report on the detection of the HNC$_3$ species for the first time 
in a prestellar core using the ASAI IRAM-30m spectral survey. We also report the detection of 4 transitions of HCCNC, one of them being already detected in the same source by \citet{jimenez-serra2016}. The first astronomical detections of HCCNC and HNC$_3$ were reported by \citet{kawaguchi1992a} and \citet{kawaguchi1992b}, respectively, towards a cold, dark molecular cloud, TMC-1. These molecules are linked to the chemistry of HC$_3$NH$^+$, previously 
detected in the ASAI observations. The observed line frequencies associated with HNC$_3$ were slightly off ($\sim$ 0.2 MHz) 
compared to the theoretical prediction given by the JPL catalog \citep{pickett1998}. We present in this letter the analysis 
of HNC$_3$ published microwave frequencies with millimeter data, up to 300 GHz.

\section{Laboratory spectroscopic considerations}

Rest frequency predictions for the linear isocyanoacetylene (HCCNC) molecule were taken from 
the JPL catalog. The entry is based on Fourier transform microwave 
(FTMW) and millimeter-wave (mmW) measurements \citep{kruger1991,guarnieri1992}, 
the dipole moment was determined by \citet{kruger1993}. 

Iminopropadienylidene (HNC$_3$) is a quasi-linear molecule, i.e., an asymmetric top rotor with a 
low barrier to linearity \citep{botschwina2003} 
such that rotational levels with $K_a \ge 1$ will be difficult to be observed in the laboratory, even 
less so in the ISM. The initial predictions for HNC$_3$ were also taken from the JPL catalog. 
The entry was based on FTMW data \citep{hirahara1993} and additional frequencies from the 
radio astronomical observations \citep{kawaguchi1992b}. 
We were able to tentatively assign emission lines for this 
species in our L1544 ASAI spectral survey. However, the predictions deviated significantly (between about 170 and 280~kHz) 
from our observed lines because the data employed for the initial predictions extended only to 47~GHz. 
\citet{kawaguchi1992b} mentioned mmW laboratory measurements of HNC$_3$ and published resulting 
spectroscopic parameters, but not the rest frequencies with their uncertainties. In the present study we 
combine FTMW data \citep{hirahara1993} displaying $^{14}$N hyperfine structure (HFS) splitting with 
the mmW laboratory data without HFS splitting to obtain improved spectroscopic parameters; 
one poorly fitting weak FTMW transition frequency was omitted. The resulting parameters are shown 
in Table \ref{spec-parameter}. Table A1 (online supplementary material) summarized experimental lines with 
uncertainties and residuals between measured rest-frequency and that calculated from the 
final spectroscopic parameters along with predictions up to 50~GHz under consideration of HFS 
splitting. Table A2 presents the corresponding data without HFS splitting up 
to 300~GHz. A dipole moment of 5.665~D was calculated by \citet{botschwina1992}. The 
predictions and associated files will be available in the CDMS \citep{endres2016}.

%%%%%%%%%%%%%%%%%%%%%%%%%%%%%%%%%%%%%%%%%%%%%%%%%%%%%%%%%%%%%%%%%%%%%%%%%%%%%%%%%%%%%%%%%%
%%%%%%%%%%%%%%%%%%%%%%%%%%%%%%%%%%%%%%%%%%%%%%%%%%%%%%%%%%%%%%%%%%%%%%%%%%%%%%%%%%%%%%%%%%

\begin{table}
\begin{center}
\caption{Spectroscopic parameters (MHz) of iminopropadienylidene (HNC$_3$). Numbers in parentheses are one standard deviation in units of the least significant digits.}
\label{spec-parameter}
\begin{tabular}[t]{lr@{}l}
\hline \hline
Parameter        & \multicolumn{2}{c}{Value} \\
\hline
$B$              & 4668&.33576~(39) \\
$D \times 10^6$  &  618&.57~(28)    \\
$eQq$            &    1&.0875~(64)  \\

\hline
\end{tabular}\\[2pt]
\end{center}
\end{table}

%%%%%%%%%%%%%%%%%%%%%%%%%%%%%%%%%%%%%%%%%%%%%%%%%%%%%%%%%%%%%%%%%%%%%%%%%%%%%%%%%%%%%%%%%%
%%%%%%%%%%%%%%%%%%%%%%%%%%%%%%%%%%%%%%%%%%%%%%%%%%%%%%%%%%%%%%%%%%%%%%%%%%%%%%%%%%%%%%%%%%

\section{Observations and results}

\begin{figure}
\begin{center}
\includegraphics[width=6.5cm,angle=0]{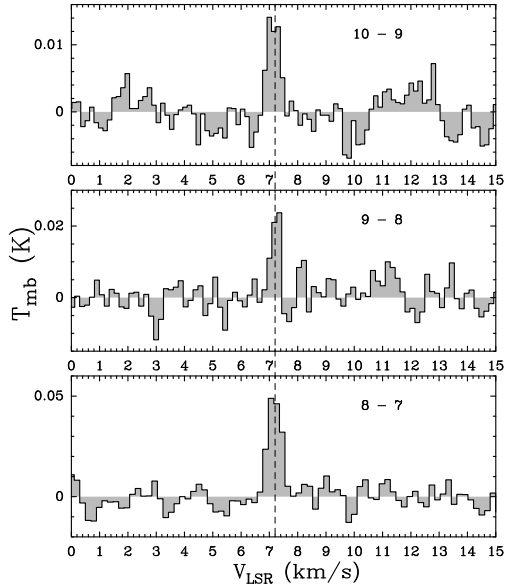}
\caption{Three detected transitions of HNC$_3$ towards L1544. The vertical line show a V$_{LSR}$ = 7.2 km~s$^{-1}$. }
\label{hnc3}
\end{center}
\end{figure}

\begin{figure}
\begin{center}
\includegraphics[width=9cm,angle=-90]{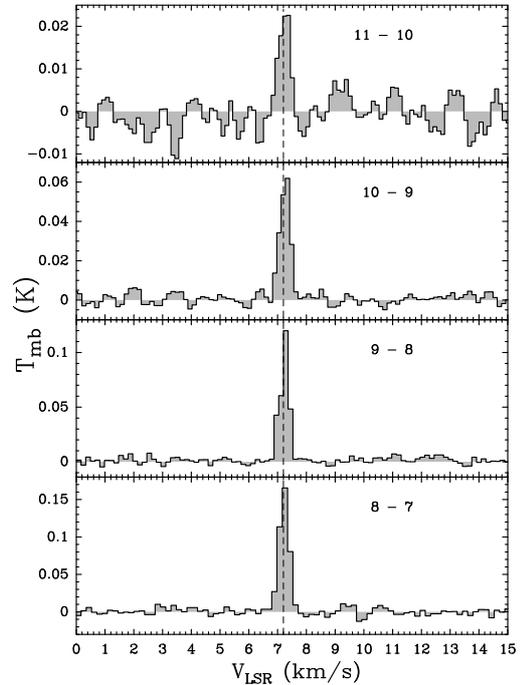}
\caption{Four detected transitions of HCCNC towards L1544. The vertical line show a V$_{LSR}$ = 7.2 km~s$^{-1}$. }
\label{hccnc}
\end{center}
\end{figure}

All the informations concerning the observations performed at the IRAM 30m can be found in \citet{vastel2014} and \citet{quenard2017}. Table \ref{lines} reports the spectroscopic parameters and the properties of the detected lines, obtained by a gaussian fitting function. For the line identification, we used the CASSIS\footnote{\url{http://cassis.irap.omp.eu}} software \citep{vastel2015a}. We present in Figure \ref{hnc3} the three HNC$_3$ transitions and in Figure \ref{hccnc} the four HCCNC transitions, detected in the ASAI spectral survey. Note that the 10--9 HCCNC transition has also been detected by \citet{jimenez-serra2016}. Both species, as well as HC$_3$N already presented in \citet{quenard2017} are produced through the HC$_3$NH$^+$ recombination reaction. This ion has previously been detected in L1544 by \citet{quenard2017}. Local thermodynamic equilibrium (LTE) analysis of HC$_3$NH$^+$ leads to a derived column density of (1--2) $\times$ 10$^{11}$ cm$^{-2}$. In the case of HC$_3$N a non-LTE analysis leads to the computation of a column density of (1.6--2.4) $\times$ 10$^{13}$ cm$^{-2}$. 

\begin{table*} 
%\tiny
\centering
\caption{Properties of the observed  HNC$_3$ and HCCNC (E$_{\textrm{up}}$ $\le$ 30\,K, A$_{\textrm{ul}}$ $\ge$ 5 $\times$ 10$^{-6}$\,s$^{-1}$) lines. Note that the errors in the integrated intensities include the statistical uncertainties from the gaussian fits of the lines. The rms has been computed over 20\,km\,s$^{-1}$.  \label{lines}}
\begin{tabular}{cccccccc}
&&&HNC$_3$&&&&\\
\hline
Transitions	& Rest Frequency & E$_{\textrm{up}}$ & A$_{\textrm{ul}}$ & FWHM            &$\int T_{\textrm{mb}}dV$ &rms \\
($J' - J''$)	&  (MHz)         &   (K)             &     (s$^{-1}$)               & (km\,s$^{-1}$)  &(mK\,km\,s$^{-1}$)       & (mK) \\
\hline
$8 - 7$     & 74692.1053     & 16.13             & 7.32 $\times$ 10$^{-5}$                        & 0.57 $\pm$ 0.05 & 31.8 $\pm$ 5.6          & 4.8\\
$9 - 8$     & 84028.2399     & 20.16             & 1.05 $\times$ 10$^{-4}$                        & 0.34 $\pm$ 0.08 & 9.5 $\pm$ 4             & 4.5\\
$10 - 9$    & 93364.2409     & 24.64             & 1.45 $\times$ 10$^{-4}$                        & 0.51 $\pm$ 0.07 & 7.6 $\pm$ 2             & 2.7\\
\hline
 &&&HCCNC&&&&\\
 \hline
 $8 - 7$  & 79484.1277 &17.17  &2.36 $\times$ 10$^{-5}$ & 0.41 $\pm$ 0.01  & 73.4 $\pm$ 2.3  & 4.3\\
 $9 - 8$   & 89419.2601 &21.46  &3.34 $\times$ 10$^{-5}$ & 0.37 $\pm$ 0.04  & 44.0 $\pm$ 9.1  & 2.7\\
 $10 - 9$    & 99354.2570 &26.23  &4.62 $\times$ 10$^{-5}$  & 0.44 $\pm$ 0.03  & 29.2 $\pm$ 4.1  & 2.5\\
$11 - 10$   & 109289.1036 &31.47  &6.19 $\times$ 10$^{-5}$  & 0.45 $\pm$ 0.07  & 11.7 $\pm$ 3.5  & 3.8\\
\hline
\end{tabular}
\end{table*}

The vertical dashed line in Figure \ref{hccnc} represents the velocity in the Local Standard of Rest (V$_{LSR}$) of 7.2 km~s$^{-1}$ for the HCCNC species showing a good match (within 0.1 km~s$^{-1}$) between the observed frequencies and the calculated frequencies quoted by the JPL archive database \citep{pickett1998}. However, the HNC$_3$ fitted frequencies from the JPL database cannot be reproduced with the L1544 V$_{LSR}$, with an observed shift of about 1 km~s$^{-1}$. We therefore use the frequencies as computed in Section 2 (see Table A2). In the absence of collisional coefficients for HNC$_3$ and HCCNC, LTE analysis has been performed using CASSIS, assuming no beam dilution. Note that LTE and non-LTE analysis for HC$_3$N led similar results and we can only assume it will be the same for HNC$_3$ and HCCNC. The best LTE fit is obtained for a low excitation temperature of (6-8) K with column densities of (0.75--2) $\times$ 10$^{11}$ cm$^{-2}$ for HNC$_3$ and (0.85--2.2) $\times$ 10$^{12}$ cm$^{-2}$ for HCCNC. 
\citet{jimenez-serra2016} estimated an observed column density of (0.3--3.0) $\times$ 10$^{12}$ cm$^{-2}$ in a (5-10) K range using one transition only. 
In any case, the observed HCCNC column density is $\sim$ 10 times higher than HNC$_3$. \\

\section{Chemical modelling}

\begin{figure*}
\includegraphics[width=0.445\hsize]{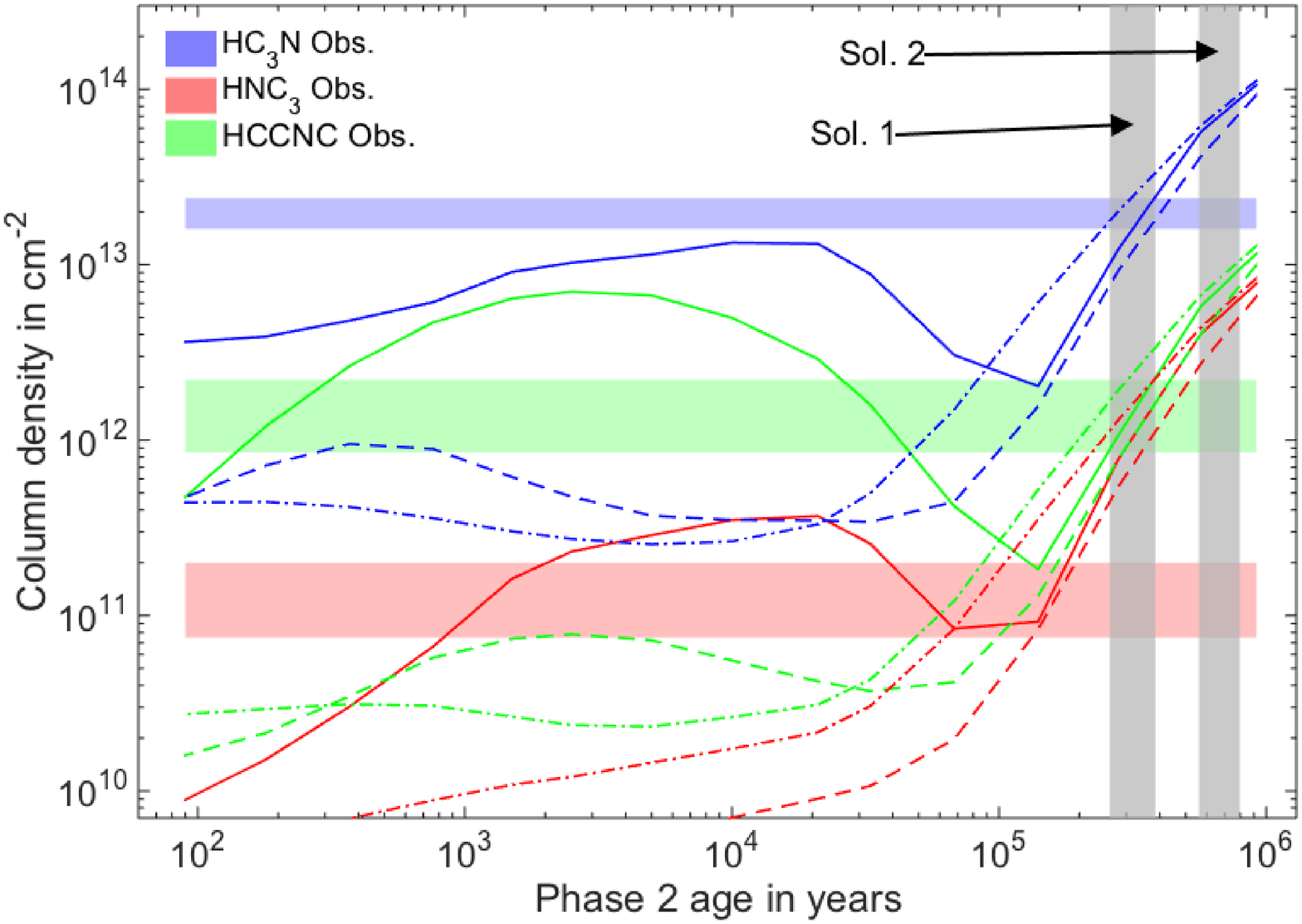}
\includegraphics[width=0.445\hsize]{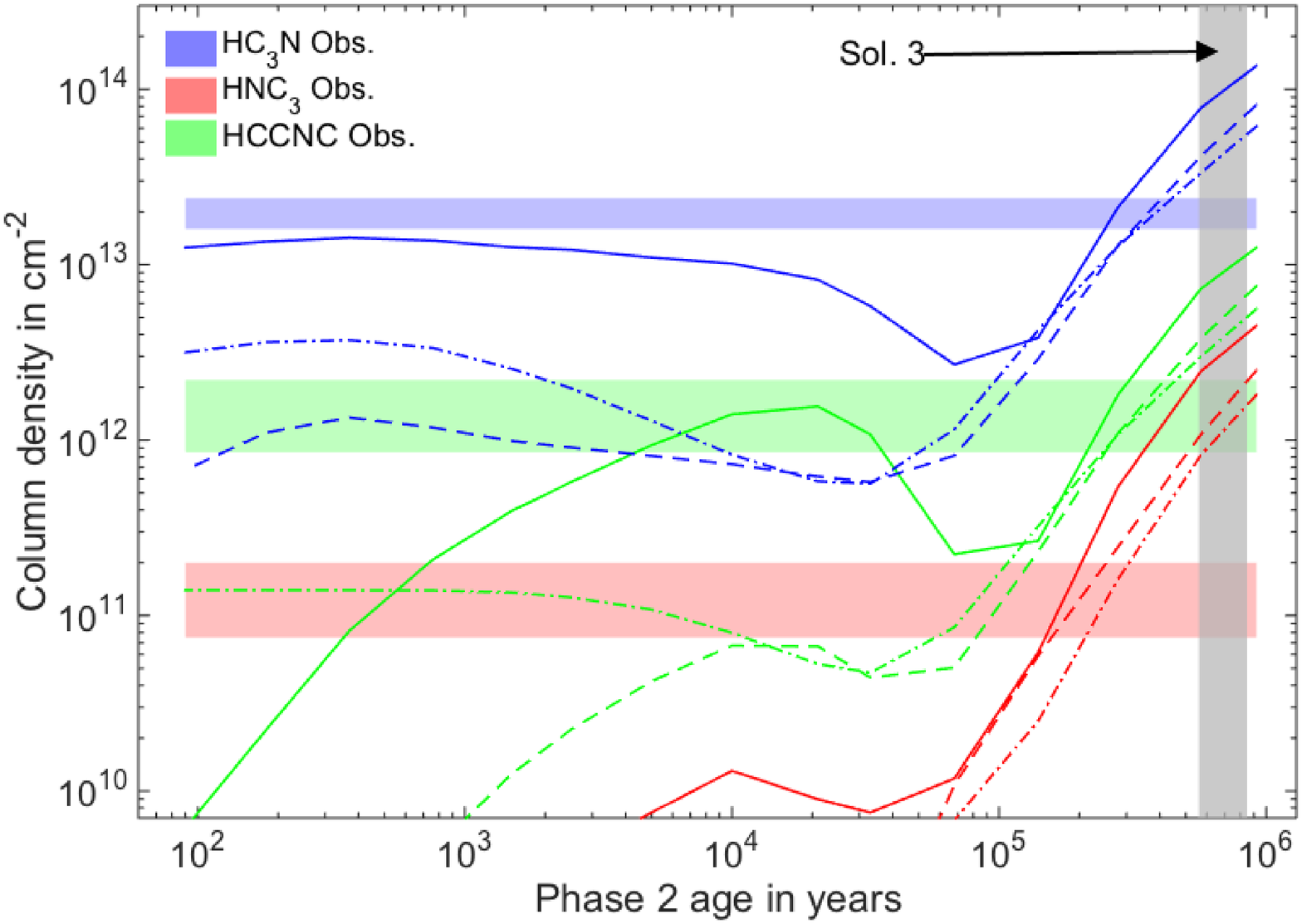}
\caption{Column densities for HC$_3$N, HNC$_3$ and HCCNC as a function of the age of the phase 2 for different initial gas densities of the ambient cloud phase: 10$^2$ cm$^{-3}$ (full line), 3 $\times$ 10$^3$ cm$^{-3}$ (dashed line), and 2 $\times$ 10$^4$ cm$^{-3}$ (dash-dotted line). The area of confidence of the observed column densities for these species is also shown. Grey areas shows the timespan area of confidence of each model based on the observed HCNH$^+$. These areas are labelled solutions 1, 2 and 3 \citep{quenard2017}. \textit{Left panel:} Model EA1 of initial atomic abundances ("low-metal abundances"). \textit{Right panel:} Model EA2 of initial atomic abundances ("high-metal abundances"). All the informations on the modelling can be found in \citet{quenard2017}.}	
\label{model}
\end{figure*}

Using a gas-grain chemical code \citep[Nautilus, described in][]{ruaud2016}, \citet{quenard2017} calculated the predicted abundance using a two-step model by varying the initial gas densities of the initial phase. The resulting abundances have been compared to the observed column densities and models with an age $\ge$ 10$^5$ years give a satisfactory agreement to the observations. The same modelling shows that HC$_3$NH$^+$, as well as HC$_3$N and HNC$_3$ are emitted mostly in an external layer at about 10$^4$ au from the center, with a temperature of about 10 K and a density of $\sim$ 10$^4$ cm$^{-3}$. 
When HC$_3$NH$^+$ recombines with an electron, the neutralized HC$_3$NH molecule has a substantial internal energy, leading to the breaking of chemical bonds. HC$_3$N, HNC$_3$ and HCCNC are all products of the HC$_3$NH$^+$ dissociative recombination (DR) reaction: 
\small
\begin{eqnarray}
	\mathrm{HC_3NH^+ + \textit{e}^-} &\longrightarrow& \mathrm{HC_3N+H} \hspace{1cm} \\
	&\longrightarrow& \mathrm{HNC_3+H}   \hspace{1cm}\\
	&\longrightarrow& \mathrm{HCCNC+H}  \hspace{0.75cm}\\
	&\longrightarrow& \mathrm{HCNCC+H}  \hspace{0.75cm}\\
	&\longrightarrow& \mathrm{HCN + C_2H}  \hspace{0.75cm}\\
	&\longrightarrow& \mathrm{HNC + C_2H}  \hspace{0.75cm}
\end{eqnarray}
\normalsize
Reactions (5) and (6) are not the main routes for the production of HCN and HNC, which are also products of the DR of HCNH$^+$. Also, the DR of HC$_3$NH$^+$ is not the main route for the production of HC$_3$N. In their theoretical analysis, \citet{osamura1999} have chosen equal branching fractions for reactions (1) and (2) that are 5 times greater than those for the products in reactions (3) and (4). They also concluded that a variation from 3 to 9 (instead of 5) do not have a major effect on the calculated abundances of the metastable isomers of HC$_3$N. The branching ratio for the DR of DC$_3$ND$^+$ have been estimated using the heavy ion storage ring CRYRING in Stockholm, Sweden \citep{vigren2012} which would, in analogy with HC$_3$NH$^+$ lead to the following branching fractions: 22\%, 22\%, 4\%, 4\%, 24\%, 24\% for reactions (1), (2), (3), (4), (5) and (6) respectively. The thermal rate coefficients for the DR of the DC$_3$ND$^+$ has been measured by  \citet{geppert2004}: 1.5 $\times$ 10$^{-6}$(T/300)$^{-0.58}$ cm$^3$~s$^{-1}$. We used the KIDA database (http://kida.obs.u-bordeaux1.fr/) and modified the HC$_3$NH$^+$ dissociative reaction rate according to \citet{geppert2004}, using the branching ratio following \citet{vigren2012}. \\
Figure B1 (online supplementary material) shows the results of the chemical modelling for the HCNH$^+$ and HC$_3$NH$^+$ ions using the modified network described above as was performed in \citet{quenard2017}. The model is divided into two phases. The first phase corresponds to the evolution of the chemistry in a diffuse or molecular cloud (hereafter ambient cloud) and depends on the assumed initial H density. The chemical composition then evolves for 10$^6$ years. The abundances from this first step are then used as initial abundances for the second step where we consider the physical structure of the L1544 prestellar core \citep{keto2014}. The HCNH$^+$ and HC$_3$NH$^+$ species are emitted from the outer layer when non-thermal desorption of other species was previously observed \citep{vastel2014}. Solutions 1, 2 and 3 (grey box in Fig. B1) represent the best comparison with the observation for the HCNH$^+$ ion (see Table 2--4 from \citet{quenard2017}). The present modelling shows that the HC$_3$NH$^+$ observations are better reproduced with the new recombination rate, compared to \citet{quenard2017}, by a factor of $\sim$ 10. We also present in Fig. \ref{model} the comparison between the modelling and the observed column densities of HCCNC, HC$_3$N and HNC$_3$. The HCCNC modelling is compatible with solution 1 which represent the low-metal initial abundances while HNC$_3$ is higher by a factor 5-10 compared to the observations. This solution has already been obtained for HC$_3$NH$^+$, HC$_3$N as well as other nitrogen species which have already been published by \citet{quenard2017}. We want to emphasise that, in the present network, the production of HNC$_3$ is dominated by the HC$_3$NH$^+$ DR while the production of HC$_3$N is dominated by the following reactions: C$_2$H$_2$ + CN and CH$_2$CCH + N. The network seems to overproduce HNC$_3$ compared to the observations. We present in Fig. B2 the comparison for the HNC$_3$ species between the chemical modelling presented in \citet{quenard2017} and the present modified network. The destruction pathway for HNC$_3$ is dominated by the electron recombination for which the rate is only estimated following the experimental dissociative attachment of HNO$_3$ \citep{adams1986,harada2008}. Note also that the branching ratios for reactions (1)--(6) are highly uncertain, so a variation could have an impact on the chemical modelling of the aforementioned species. Note also that the HNC$_3$, HC$_3$N and HCCNC molecules may be produced from the linear HCCNCH$^+$ ion for which we have no spectroscopic information so far.  
 
\section{Comparison with the TMC-1 dark cloud}

HC$_3$NH$^+$ (2 transitions), HNC$_3$ (3 transitions) and HCCNC (3 transitions) have been detected for the first time in the cold starless cloud TMC-1 (position of maximum cyanopolyyne emission: $\alpha(2000)=04^h41^m42.47^s, \delta(2000)=+25^{o}41^{\prime}27.1^{\prime\prime}$) using the Nobeyama 45~m telescope \citep{kawaguchi1992a,kawaguchi1992b,kawaguchi1994}. Table \ref{comparison} shows the comparison between the TMC-1 dark cloud and the L1544 prestellar core for the observations of these related species. The TMC-1 dark cloud is at an earlier stage than the L1544 prestellar core which is believed to be on the verge of collapse. It is important to note that, for TMC-1, the reported derived excitation temperatures for HC$_3$NH$^+$, HNC$_3$ and HCCNC are (6.5 $\pm$ 0.1) K, (5.0 $\pm$ 0.7) K and (6.4 $\pm$ 0.6) K, respectively, at the lower end of the (6-8) K temperature range that we derived from the L1544 observations. The detection of the HC$_3$N species has been reported for the first time in TMC-1 by \citet{suzuki1992} who reported the column density to be 1.71 $\times$ 10$^{14}$ cm$^{-2}$ using the J = 5--4 transition with the NRO 45-m telescope, if the excitation temperature is 6.5 K.  \citet{takano1998} then reported the analysis of  the J = 2--1, 4--3, 5--4, and 10--9 transitions of HC$_3$N at the 45-m Nobeyama telescope and derived a rotational temperature of (7.1 $\pm$ 0.1) K and column density of (1.6 $\pm$ 0.1) $\times$ 10$^{14}$ cm$^{-2}$. 
The HC$_3$N/HC$_3$NH$^+$, HNC$_3$/HC$_3$NH$^+$ and HCCNC/HC$_3$NH$^+$ ratios in both sources are then similar, reflecting the cold and dense conditions in both environments: 160 $\pm$ 42, 0.4 $\pm$ 0.1 and 2.9 $\pm$ 0.8 respectively for TMC-1, 80--240, 0.38--2, 4.25--22 respectively for L1544. 

\begin{table}
%\tiny
\centering
\caption{Comparison of the observed column densities (in cm$^{-2}$) between the TMC-1 dark cloud and the L1544 prestellar core. \label{comparison}}
\begin{tabular}{ccc}
 	                        & TMC-1	&      L1544	\\
\hline
HC$_3$N               &  (1.6 $\pm$ 0.1) $\times$ 10$^{14}$    &   (1.6--2.4) $\times$ 10$^{13}$   \\
                               & \citet{takano1998}                  &   \citet{quenard2017}\\
HNC$_3$               &  (3.8 $\pm$ 0.6) $\times$ 10$^{11}$    &   (0.75--2) $\times$ 10$^{11}$   \\
                               & \citet{kawaguchi1992a}                  &   this paper\\
HCCNC                  &  (2.9 $\pm$ 0.2) $\times$ 10$^{12}$    &   (0.85--2.2) $\times$ 10$^{12}$\\
                               & \citet{kawaguchi1992b}                  &   this paper\\
HC$_3$NH$^+$     &(1.0 $\pm$ 0.2) $\times$ 10$^{12}$    &   (1--2) $\times$ 10$^{11}$ \\
                               & \citet{kawaguchi1994}                  &   \citet{quenard2017}\\
\end{tabular}
\end{table} 

\section{Conclusions}
HNC$_3$ has been detected for the first time in a prestellar core, L1544. We computed its line frequencies, necessary for the line identification and estimation of the column density. We used a gas-grain chemical modelling and compared all related nitrogen species with the observations. Comparison of the observations with the model predictions suggests that the emission from HC$_3$NH$^+$, HC$_3$N, HNC$_3$ and HCCNC originates in the outer layer where non-thermal desorption of other species was previously observed.  We conclude that the modelled abundances are consistent with the observations, except for HNC$_3$, considering a late stage of the evolution of the prestellar core, compatible with previous observations.

\section*{Acknowledgements}
\footnotesize{Based on observations carried out with the IRAM 30 m Telescope. IRAM is supported by INSU/CNRS (France), MPG (Germany), and IGN (Spain). K.K. and M.O. are very grateful to S. Yamamoto and S. Saito for the opportunity to carry out laboratory spectroscopic investigations on HNC$_3$ at the Nagoya University. H.S.P.M. thanks Y. Endo for comments on the FTMW data of HNC$_3$. M.O. is supported by the JSPS Kakenhi grant number JP15H03646. C.V. would like to thank J.-C. Loison for a fruitful discussion on the reaction rates and branching fraction of the DR of the HC$_3$NH$^+$ ion. D.Q. acknowledges the financial support received from the STFC through an Ernest Rutherford Grant (proposal number ST/M004139).}

%%%%%%%%%%%%%%%%%%%%%%%%%%%%%%%%%%%%%%%%%%%%%%%%%%



\bibliographystyle{mnras}
%\bibliography{/Users/vastel/ARTICLES/All_ref} % if your bibtex file is called example.bib
\bibliography{mnras_L1544_V3_withappendix.bbl}
%%%%%%%%%%%%%%%%%%%%%%%%%%%%%%%%%%%%%%%%%%%%%%%%%%
%%%%%%%%%%%%%%%%% APPENDICES %%%%%%%%%%%%%%%%%%%%%
\begin{appendix}

\appendix
\section{Supplementary tables}

\begin{table}
%\begin{center}
\caption{Quantum numbers $J$ and $F$ and observed or calculated$^a$ 
transition frequency (MHz) of iminopropadienylidene, HNC$_3$, up to 50~GHz 
with consideration of $^{14}$N HFS, their uncertainty$^b$ Unc. (in kHz) and 
residual O$-$C (in kHz) between observed frequency and that calculated from 
the final set of spectroscopic parameters.}
\label{HFS-lines}
\begin{tabular}{ccr@{}lr@{}lr@{}l}
\hline \hline
$J' - J''$ & $F' - F''$ & \multicolumn{2}{c}{Frequency} 
& \multicolumn{2}{c}{Unc.} & \multicolumn{2}{c}{O$-$C} \\
\hline
 1 $-$ 0 & 0 $-$ 1 &   9336&.130    & 5&.  &    4&.71 \\
 1 $-$ 0 & 2 $-$ 1 &   9336&.616    & 5&.  &    1&.33 \\
 1 $-$ 0 & 1 $-$ 1 &   9336&.945    & 5&.  &    4&.08 \\
 2 $-$ 1 & 1 $-$ 1 &  18672&.778    & 5&.  & $-$1&.49 \\
 2 $-$ 1 & 1 $-$ 2 &  18673&.1057   & 1&.8 &     & $^d$   \\
 2 $-$ 1 & 3 $-$ 2 &  18673&.303    & 5&.  &    3&.06 \\
 2 $-$ 1 & 2 $-$ 1 &  18673&.3232   & 1&.5 &     &  $^d$  \\
 2 $-$ 1 & 1 $-$ 0 &  18673&.598    & 5&.  &    2&.88 \\
 2 $-$ 1 & 2 $-$ 2 &  18673&.675$^c$& 5&.  &   25&.51 \\
 3 $-$ 2 & 2 $-$ 2 &  28009&.4584   & 3&.4 &     &  $^d$  \\
 3 $-$ 2 & 2 $-$ 3 &  28009&.8079   & 2&.3 &     &  $^d$  \\
 3 $-$ 2 & 4 $-$ 3 &  28009&.9348   & 2&.3 &     &   $^d$ \\
 3 $-$ 2 & 3 $-$ 2 &  28009&.9477   & 2&.3 &     &   $^d$ \\
 3 $-$ 2 & 2 $-$ 1 &  28010&.0021   & 2&.4 &     &    $^d$\\
 3 $-$ 2 & 3 $-$ 3 &  28010&.2973   & 3&.3 &     &    $^d$\\
 4 $-$ 3 & 3 $-$ 3 &  37346&.0616   & 3&.8 &     &   $^d$ \\
 4 $-$ 3 & 3 $-$ 4 &  37346&.4241   & 3&.0 &     &    $^d$\\
 4 $-$ 3 & 5 $-$ 4 &  37346&.5195   & 3&.0 &     &   $^d$ \\
 4 $-$ 3 & 4 $-$ 3 &  37346&.5277   & 3&.1 &     &    $^d$\\
 4 $-$ 3 & 3 $-$ 2 &  37346&.5510   & 3&.1 &     &    $^d$\\
 4 $-$ 3 & 4 $-$ 4 &  37346&.8902   & 4&.0 &     &    $^d$\\
 5 $-$ 4 & 4 $-$ 4 &  46682&.5952   & 4&.2 &     &    $^d$\\
 5 $-$ 4 & 6 $-$ 5 &  46683&.0426   & 3&.8 &     &    $^d$\\
 5 $-$ 4 & 5 $-$ 4 &  46683&.0483   & 3&.8 &     &    $^d$\\
 5 $-$ 4 & 4 $-$ 3 &  46683&.0612   & 3&.8 &     &    $^d$\\
 5 $-$ 4 & 5 $-$ 5 &  46683&.4190   & 4&.6 &     &    $^d$\\
\hline
\end{tabular}\\ %[2pt]
%\end{center}
$^a$ Frequencies and uncertainties with residuals O$-$C 
 are experimental ones; those without were calculated from the final 
 set of spectroscopic parameters. $^b$ 1~$\sigma$ values. 
 $^c$ Weak line, frequency omitted because of large residual. $^d$ Calculated frequency and uncertainty.
\end{table}

\begin{table}
%\begin{center}
\caption{Quantum numbers $J$ and $F$ and observed or calculated$^a$ 
transition frequency (MHz) of iminopropadienylidene, HNC$_3$, up to 
300~GHz without consideration of $^{14}$N HFS, their uncertainty$^b$ 
Unc. (in kHz) and residual O$-$C (in kHz) between observed frequency and that 
calculated from the final set of spectroscopic parameters.}
\label{nohfs-lines}
\begin{tabular}{cr@{}lr@{}lr@{}l}
\hline \hline
$J' - J''$ & \multicolumn{2}{c}{Frequency} 
& \multicolumn{2}{c}{Unc.} & \multicolumn{2}{c}{O$-$C} \\
\hline
  1 $-$  0 &    9336&.6690 &  0&.8 &      &   $^c$ \\
  2 $-$  1 &   18673&.3232 &  1&.5 &      &   $^c$ \\
  3 $-$  2 &   28009&.9477 &  2&.3 &      &   $^c$ \\
  4 $-$  3 &   37346&.5277 &  3&.1 &      &   $^c$ \\
  5 $-$  4 &   46683&.0483 &  3&.8 &      &   $^c$ \\
  6 $-$  5 &   56019&.4947 &  4&.5 &      &   $^c$ \\
  7 $-$  6 &   65355&.8519 &  5&.1 &      &   $^c$ \\
  8 $-$  7 &   74692&.1053 &  5&.7 &      &   $^c$ \\
  9 $-$  8 &   84028&.2399 &  6&.3 &      &   $^c$ \\
 10 $-$  9 &   93364&.2409 &  6&.8 &      &   $^c$ \\
 11 $-$ 10 &  102700&.0934 &  7&.2 &      &  $^c$  \\
 12 $-$ 11 &  112035&.7826 &  7&.6 &      &   $^c$ \\
 13 $-$ 12 &  121371&.2937 &  7&.9 &      &   $^c$ \\
 14 $-$ 13 &  130706&.6118 &  8&.1 &      &   $^c$ \\
 15 $-$ 14 &  140041&.7221 &  8&.3 &      &   $^c$ \\
 16 $-$ 15 &  149376&.560  & 25&.  & $-$49&.62 \\
 17 $-$ 16 &  158711&.2597 &  8&.4 &      &  $^c$  \\
 18 $-$ 17 &  168045&.656  & 25&.  &  $-$1&.31 \\
 19 $-$ 18 &  177379&.801  & 25&.  &    13&.25 \\
 20 $-$ 19 &  186713&.634  & 25&.  &  $-$2&.12 \\
 21 $-$ 20 &  196047&.188  & 25&.  &     0&.43 \\
 22 $-$ 21 &  205380&.427  & 25&.  &  $-$0&.27 \\
 23 $-$ 22 &  214713&.3404 &  7&.1 &      &  $^c$  \\
 24 $-$ 23 &  224045&.9120 &  7&.0 &      &   $^c$ \\
 25 $-$ 24 &  233378&.159  & 25&.  &    31&.65 \\
 26 $-$ 25 &  242709&.925  & 25&.  & $-$46&.55 \\
 27 $-$ 26 &  252041&.460  & 25&.  &    30&.23 \\
 28 $-$ 27 &  261372&.472  & 25&.  & $-$15&.16 \\
 29 $-$ 28 &  270703&.119  & 25&.  &  $-$9&.86 \\
 30 $-$ 29 &  280033&.334  & 25&.  &  $-$6&.05 \\
 31 $-$ 30 &  289363&.126  & 25&.  &    20&.14 \\
 32 $-$ 31 &  298692&.4115 & 15&.8 &      &  $^c$  \\
\hline
\end{tabular}\\%[2pt]
%\end{center}
$^a$ Frequencies and uncertainties with residuals O$-$C are experimental ones; those without were calculated from the final set of spectroscopic parameters. $^b$ 1~$\sigma$ values. $^c$ Calculated frequency and uncertainty.
\end{table}

\section{Supplementary Figures}

\begin{figure*}
\includegraphics[width=0.45\linewidth]{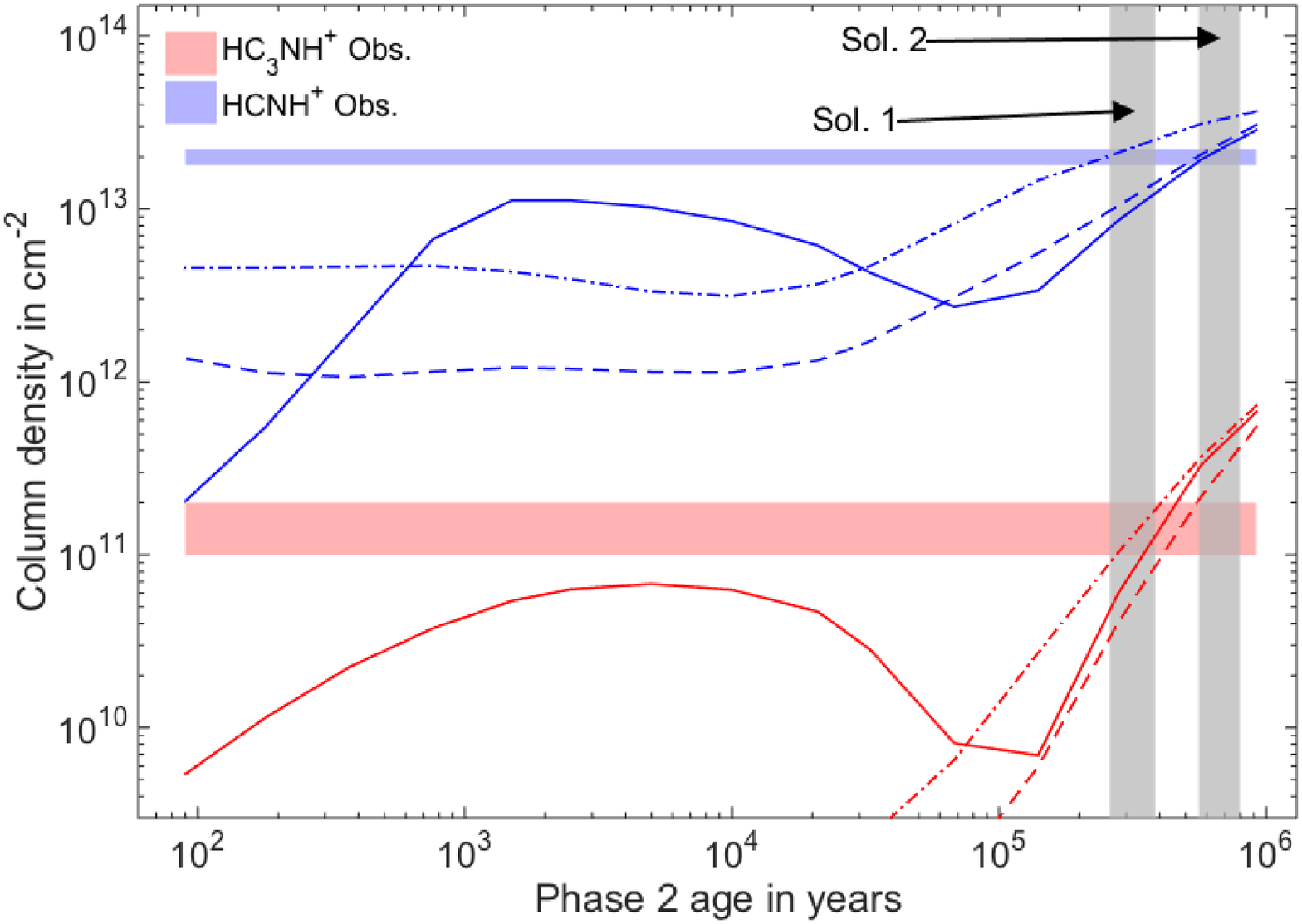}
\includegraphics[width=0.45\linewidth]{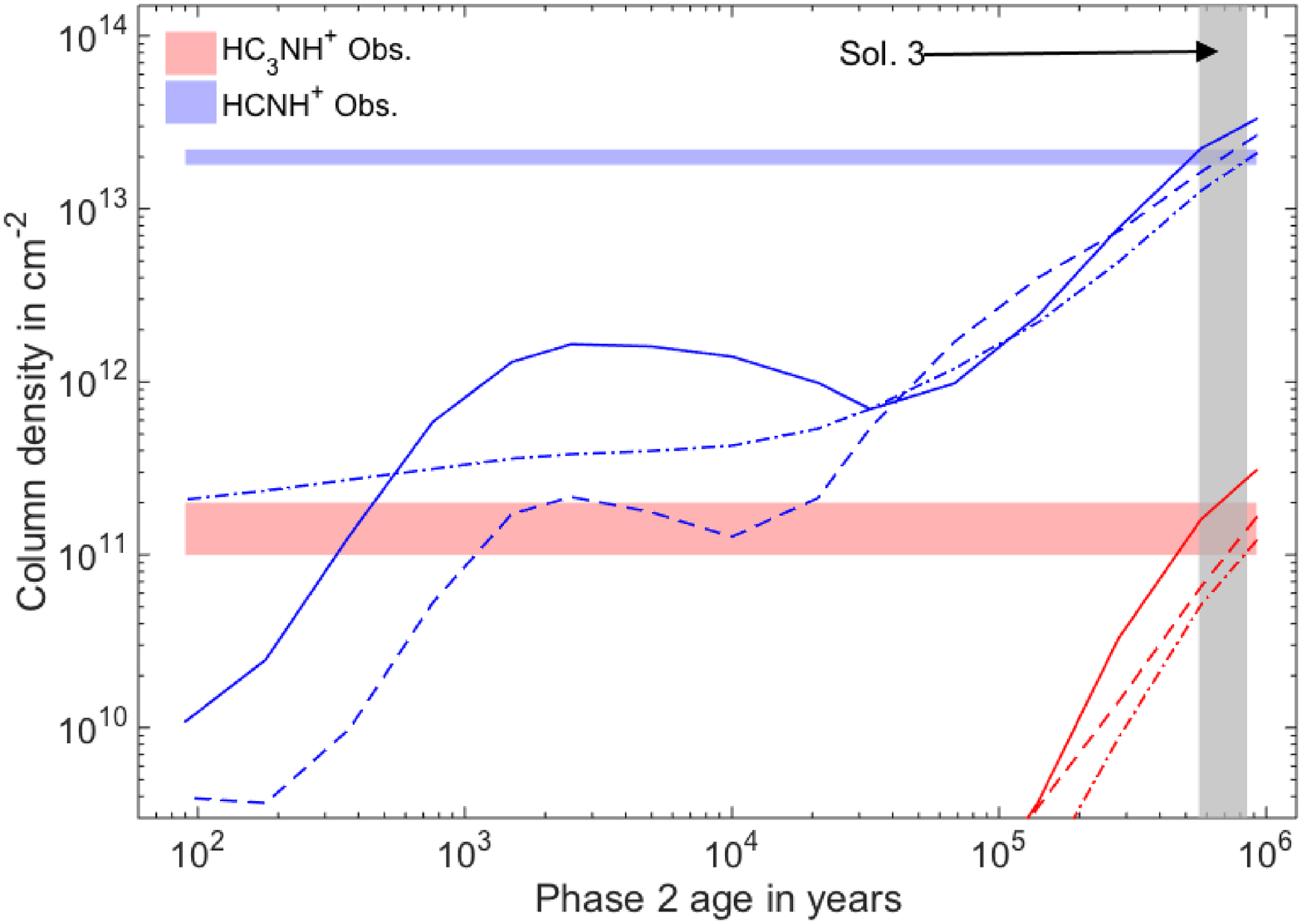}
\caption{Column densities for HC$_3$NH$^+$, and HCNH$^+$ as a function of the age of the phase 2 for different initial gas densities of the ambient cloud phase: 10$^2$ cm$^{-3}$ (full line), 3 $\times$ 10$^3$ cm$^{-3}$ (dashed line), and 2 $\times$ 10$^4$ cm$^{-3}$ (dash-dotted line). The area of confidence of the observed column densities for these species is also shown. Grey areas shows the timespan area of confidence of each model based on the observed HCNH$^+$. These areas are labelled solutions 1, 2 and 3 \citep{quenard2017}. \textit{Left panel:} Model EA1 of initial atomic abundances ("low-metal abundances"). \textit{Right panel:} Model EA2 of initial atomic abundances ("high-metal abundances"). All the informations on the modelling can be found in \citet{quenard2017}.}
\label{model_hcnhp}
\end{figure*}

\begin{figure*}
\includegraphics[width=0.45\linewidth]{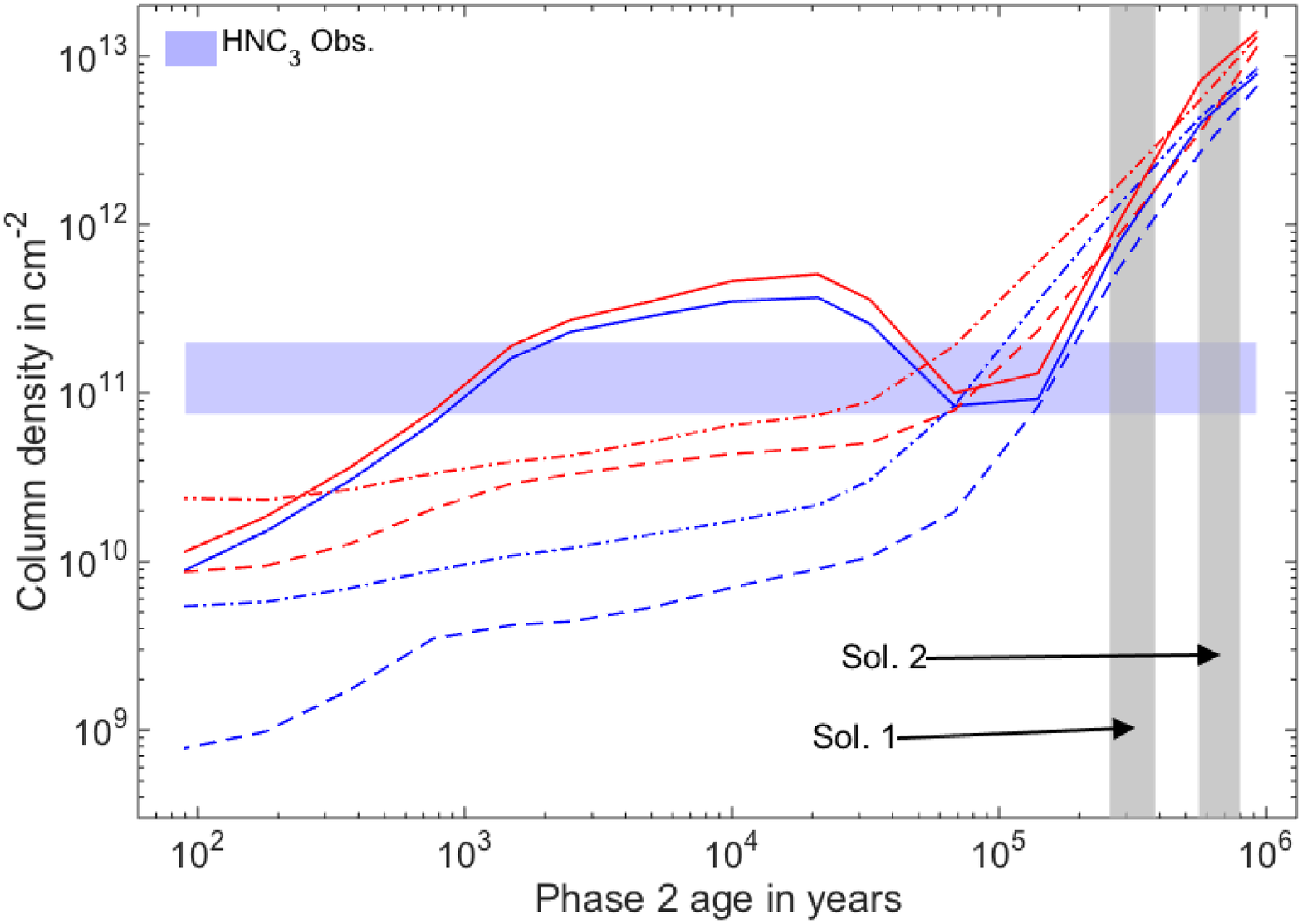}
\includegraphics[width=0.45\linewidth]{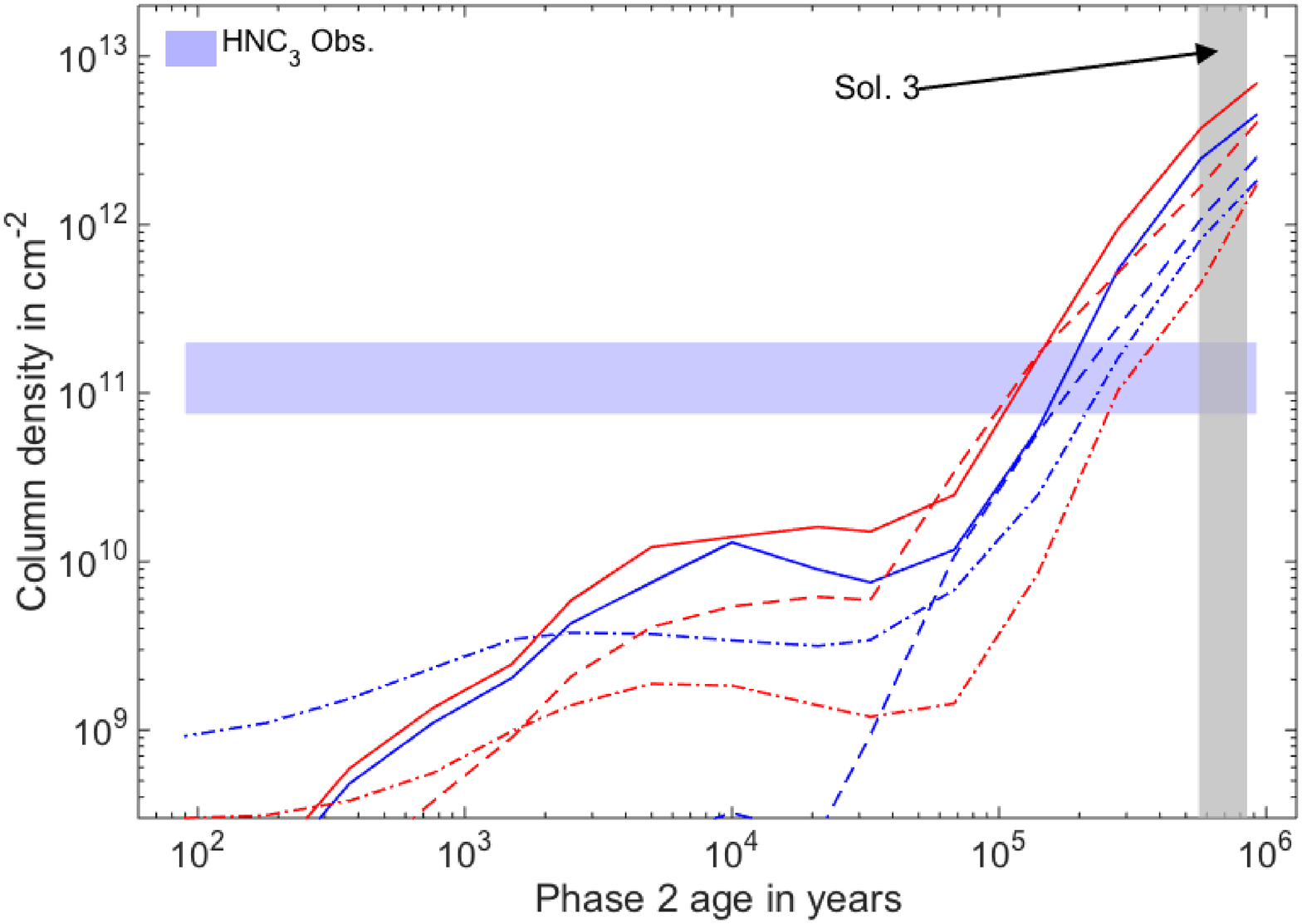}
\caption{Column densities for HNC$_3$ as a function of the age of the phase 2 for different initial gas densities of the ambient cloud phase: 10$^2$ cm$^{-3}$ (full line), 3 $\times$ 10$^3$ cm$^{-3}$ (dashed line), and 2 $\times$ 10$^4$ cm$^{-3}$ (dash-dotted line). The red lines correspond to the chemical modelling from \citet{quenard2017} while the blue lines correspond to the modified network explained in this paper. The area of confidence of the observed column densities for these species is also shown. Grey areas shows the timespan area of confidence of each model based on the observed HCNH$^+$. These areas are labelled solutions 1, 2 and 3 \citep{quenard2017}. \textit{Left panel:} Model EA1 of initial atomic abundances ("low-metal abundances"). \textit{Right panel:} Model EA2 of initial atomic abundances ("high-metal abundances"). All the informations on the modelling can be found in \citet{quenard2017}.}
\label{comp_hnc3}
\end{figure*}

\end{appendix}

%%%%%%%%%%%%%%%%%%%%%%%%%%%%%%%%%%%%%%%%%%%%%%%%%%

% Don't change these lines
\bsp	% typesetting comment
\label{lastpage}
\end{document}